\documentclass[11pt, preprint]{aastex}

\newcommand{\be}  { \begin{equation} }
\newcommand{\ee}  { \end{equation}   }
\newcommand{\bea} { \begin{eqnarray} }
\newcommand{\eea} { \end{eqnarray}   }
\newcommand{\rinf}{ \ensuremath{R_{\infty} } }

\newcommand{\rg}  { \ensuremath{R_{\rm g}  } }

\newcommand{\msun}{ \ensuremath{M_{\odot}  } }

\newcommand{\bbag}{ \ensuremath{{\bar B}   } }

\newcommand{\pdot}{ \ensuremath{\dot P     } }
\newcommand{\Edot}{ \ensuremath{\dot E     } }

\newcommand{\Phimax}{ \ensuremath{\Phi_\mathrm{max}} }

\begin{document}

\title{Is PSR B0943+10 a low-mass quark star?}

\author{Y. L. Yue$^1$, X. H. Cui$^1$, and R. X. Xu$^{2,1}$}
\affil{$^1$School of Physics, Peking University, Beijing 100871, China\\
$^2$CCAST (World Laboratory), P.O. Box 8730, Beijing 100080, China
}

\begin{abstract}
A recent X-ray observation has shown that the radio pulsar PSR
B0943+10, with clear drifting subpulses, has a much smaller polar
cap area than that of conventional pulsars with mass of
$\sim\msun$ and radius of $\sim10$ km.
\citet{zhang05} addressed then that this new result conflicts with
the standard vacuum gap model.
Nonetheless, the discrepancy could be explained if PSR B0943+10 is
actually a low-mass quark star. It is found that the potential
drop in the open-field-line region of oblique pulsars (i.e.,
inclination angle $\alpha\neq 0$) might be $\sim 10^2$ times that
of aligned pulsars, and that PSR B0943+10 with $\alpha = 12.4^{\rm
o}$ could be well above the deathline.
We thus conclude that the Ruderman-Sutherland-type vacuum gap
model still works well for this pulsar if it is a bare quark star
with a mass of $\sim 0.02M_\odot$ and a radius of $\sim 2.6$ km.
\end{abstract}

\keywords{
 dense matter ---
 pulsars: general ---
 pulsars: individual (PSR B0943+10) ---
 stars: neutron
 }

\section{Introduction}

The nature and emission mechanism  of pulsars remain as puzzles
for nearly forty years. There are various kinds of emission models
for the particle acceleration in pulsar magnetosphere, such as the
vacuum gap \citep[e.g.,][hereafter RS75]{RS75}, the
space-charge-limited flow \citep[e.g.,][]{Arons79}, the outer gap
\citep[e.g.,][]{Cheng86}, and the core and annular gaps
\citep[e.g.,][]{Qiao04}.
Due to the limits of observations and the difficulties in the
electrodynamics of pulsar magnetosphere, we still could not know
which one really works. An interesting phenomena to discriminate
these models is pulsar subpulse drifting.
RS75 vacuum gap model is by far the most successful model to
understand this drifting subpulse phenomenon. The model introduces
polar cap sparks, which demonstrate $\mathit{\mathbf{E \times B}}$
drift around the magnetic axis.
The pulsar PSR B0943+10 is one of the best-studied
subpulse-drifting pulsars. With spin period $P = 1.1\mathrm{~s}$
and period derivative $\pdot = 3.5\times10^{-15} \mathrm{~s/s}$,
PSR B0943+10 is not special on the $P$-$\pdot$ diagram. The
distance of this source is $0.63\pm0.10$ kpc, which is derived
from its dispersion measure \citep{zhang05}. By fitting its
polarization data, \citet{Lyne88} obtained an inclination angle
(angle between the spin axis and the magnetic axis) of $\alpha =
12.4^\circ$ and a view angle (angle between the spin axis and the
line-of-sight) of $\zeta = 18.2^\circ$.
\citet{Deshpande99, Deshpande01} observed the pulsar and identified
20 sparks rotating with a period of $P_3 = 37P$.

In the RS75 vacuum gap model, about half of the energetic
particles would hit the polar gap surface, so the polar gap will
be heated to emit X-rays. Thermal X-ray emission from the cap
region should then be observable.
In order to test the vacuum gap model, \citet*{zhang05} observed
PSR B0943+10 with {\em XMM-Newton}, but obtained a rather small
thermal polar cap area, $A = 10^3 [T/(3 \mathrm{~MK})]^{-4}
\mathrm{~m^2}$, where $T$ is the polar cap surface temperature
(with 1-$\sigma$ error, $A = 0.3\times10^3\mathrm{~m^2}\sim
5\times10^3\mathrm{~m^2}$, and $T = 2.0\mathrm{~MK} \sim 4.2
\mathrm{~MK}$), whereas a power-law fit is also acceptable (a
detail discussion about this issue will be presented in \S3).
Though the distance of PSR B0943+10 ($0.63\pm0.10$ kpc)
\citep{zhang05} is not very accurate, its uncertainty could not be
significant comparing to that of polar cap area. The conventional
RS75 polar cap area, $6\times10^4\mathrm{~m^2}$, would be then ten
times more that observed. Therefore, the observational
uncertainties might not be able to explain this discrepancy. To
alleviate this difficulty, \citet{zhang05} suggested a strong
multipole magnetic field in the polar cap region \citep[see
also][]{Gil05, Gil06} for this pulsar.

Though RS75 vacuum gap model with a user-friendly nature is
successful to explain subpulse drifting, there are unfortunately
two drawbacks.
One is the so-called binding energy problem. RS75 model requires
that the binding energy of the ions on the neutron star surface
should be larger than $\sim 10$ keV, which should be in doubt
\citep[e.g.,][]{xu99}.
The other is: RS75 model could not be able to apply to half of
radio pulsars, the {\em antipulsars} with
$\mathit{\mathbf{\Omega\cdot B}} > 0$. The RS75 model may works
for $\mathit{\mathbf{\Omega\cdot B}} < 0$ if ions can be bound on
the surface, but in the case of $\mathit{\mathbf{\Omega \cdot B}}
> 0$, negatively charged electrons are required to be bound on the
surface, which can hardly be attained.
A hypothetical plasma-phase-condensation (i.e., magnetic metal)
transition was suggested \citep{ls97} for neutron star atmospheres
with high field $B$ and low temperature $T$, and was applied to
interpret the binding of electrons \citep{Usov95} and the thermal
X-ray spectrum \citep{tzd04}.
Some calculations show that magnetic metal may form if
$B\gtrsim10^{13} \mathrm{~G}$ and surface temperature $T <
3.7\times 10^5 \mathrm{~K}$ \citep{Usov95}. Actually, few pulsars
can fulfill these two criteria simultaneously, so the problem for
antipulsars still exists.
In order to overcome these two difficulties, some new ideas are
proposed, e.g., non-dipolar surface magnetic fields \citep{Gil02},
a partial flow of iron ions \citep{Gil03}, and quark stars without
crusts \citep{xu99}.

Quark stars are composed of unconfined free quarks and
gluons, which were proposed soon after Gell-Mann's idea of quarks
and were studied extensively ever since \cite[e.g.,][see
\cite{Xu03} for a short review]{Ivanenko69, Itoh70, Bodmer71,
Witten84, Alcock86}. Though there are some observational hints
that pulsars may be quark stars, whether pulsars are really normal
neutron stars or quark stars is still an open question.
It was generally believed that only strange stars with crusts
(mass $\sim 10^{-5}M_\odot$), being similar to the outer parts of
neutron stars, could work as radio pulsars, because a bare quark
surface can not supply particles to form pulsar's magnetospheres
and a bare strange star may readily accrete matter from its
``dirty'' environment \citep{Alcock86}.
However, this view was re-considered by \cite{xq98}, who addressed
that {\em bare} strange stars can also have magnetospheres and
thus radiate radio waves, since the vacuum in strong
electromagnetic static fields just above the quark surfaces
is unstable and would create $e^\pm$ pairs\footnote{%
This is the condition that the RS75-type vacuum gap model requires.
} %
although no charged particles can be pulled out from the surfaces.
Particularly, \citet{xu99} have argued then that PSR B0943+10 is
a bare strange quark star rather than a normal neutron star, which
solves the binding energy problem as well as the antipulsar
problem in a very simple way.
It was also found that, due to rapid rotation and strong
magnetic field, a crust could hardly form even in an accretion
phase of a bare strange star unless the accretion rate is much
higher than the Eddington rate \citep{xuzq01,xu02}.

In this letter, we investigated PSR B0943+10 with the new
observation by \cite{zhang05} under this quark star model.
In the model, the polar cap area problem would be
solved naturally because a quark star could have a low mass, a
small radius \citep{Alcock86}, and consequently a small polar cap
area. The binding energy problem would not be faced under the bare
quark star model since the binding energy of quarks and electrons
on the quark surfaces should be high enough for RS75-type vacuum
gaps to work in both cases of $\mathit{\mathbf{ \Omega \cdot B}} >
0$ and $\mathit{\mathbf{ \Omega \cdot B}} < 0$
\citep{xu99,xuzq01}.

\section{The Model}

For an aligned pulsar, as assumed in RS75, the polar cap radius is
$ r_\mathrm{pc} = (2\pi R^3/cP)^{1/2}$, where $R$ is the star
radius and $c$ is the light speed. The polar cap area of PSR
B0943+10 is $A_\mathrm{pc} = \pi r_\mathrm{pc}^2 \simeq
6\times10^4 \mathrm{~m^2}$ when using $R$ = 10 km and $P$ = 1.1 s.
This is much larger than the observational result of $\sim 10^3
\mathrm{~m^2}$ \citep{zhang05}.
If PSR B0943+10 is a quark star, the above problem would be solved
because a quark star could have a low mass and a small radius
\citep{Alcock86}. Applying $A_\mathrm{pc} \simeq 10^3
\mathrm{~m^2}$, one obtains that the star radius to fit the
observation is only 2.6 km.
The internal density of a low-mass quark star is almost
homogeneous in this case. The star's mass can be well approximated
by $M \simeq (4/3)\pi R^3 \rho$, where $\rho$ is the density of
the quark star.
The density $\rho$ could be a few times of the saturation nuclear
density $\rho_\mathrm{n}$. The exact value of $\rho$ is not clear
since no strong constrain has been obtained from experiments or
observations. Using a typical density $\rho = 2 \rho_\mathrm{n}$,
we can get the star's mass $M \simeq 0.019\msun$.
It is actually unclear what the phase-transition density is in the
regime of high-density but low-temperature, although lattice
quantum chromodynamics (QCD) shows the critical temperature is
($150\sim 200$) MeV for temperature-dominated case. Nevertheless,
the critical density could be only about $2\rho_\mathrm{n}$ if
neutrons (and protons) keep about 1 fm in radius. The average
density of low-mass strange quark stars could be about $4\bbag
\sim (4.4\sim 8.0) \times 10^{14} \mathrm{~g~cm^{-3}}$, where the
bag constant $\bbag$ is reasonably $(60\sim 110)
\mathrm{~MeV~fm^3}$. Even for an extreme case of $\bbag = 250
\mathrm{~MeV~fm^3}$, we have $\rho = 1.8\times10^{15}
\mathrm{~g~cm^{-3}}$ ($\sim 6.7\rho_\mathrm{n}$), $M = 0.06
\msun$, and the redshift factor $\rinf/R = (1-\rg/R)^{1/2} =
1.04$, where $\rg = 2GM/c^2$. This shows that the general
relativistic effect correction is very small and can be omitted
for low mass quark stars.

An effective electric force to power a pulsar in RS75 model
results from a potential drop between the magnetic axis (also the
spin axis for aligned pulsars) and the last open field line. The
potential drop is
$\Phi = \Omega B R^2\sin^2\theta/(2c)$,
where $\Omega = 2\pi/P$, $B$ is the star surface magnetic field
strength at the magnetic polar, $\sin\theta = r_\mathrm{pc}/R =
({R}/{R_\mathrm{LC}})^{1/2}$, $R_\mathrm{LC} = c/\Omega$ is the
light cylinder radius, and $\theta$ is half the opening angle of
the polar cap. We have effectively,
$\Phi \simeq 3\times10^{16}R_6^2B_{12}P\sin^2\theta/P$ Volts,
where $R_6 = R/(10^6 \mathrm{~ cm})$, $B_{12} = B/(10^{12}
\mathrm{~G})$, and $P$ in units of $1 \mathrm{~s}$.
One can get $\Phi = 6.6\times 10^{11} \mathrm{~V} < \Phi_\mathrm{c}
\sim 10^{12} \mathrm{~V}$ for PSR B0943+10 if it has $M = 0.019\msun$
and $R = 2.6 \mathrm{~km}$,
where $\Phi_\mathrm{c}$ is the critical
voltage which is obtained by requiring all the observed radio
pulsars have $\Phi > \Phi_\mathrm{c}$.
The field $B=6.8\times 10^{12}$ G is obtained from
Eq.(\ref{eq:B}).
If $\Phi < \Phi_\mathrm{c}$, particles would not be accelerated to
have enough energy to form sparks, and thus the star should not
give out radio emission. Actually, the death-line criteria is not
very certain yet, so we use a generally accepted constant
potential drop of $\Phi_{\rm c} \sim 10^{12} \mathrm{~V}$ for the
sake of simplicity. Since the criteria is not strict, a quark star
with a potential drop of $6.6\times10^{11} \mathrm{~V}$ might be
able to give out radio emission.

Nevertheless, the following calculation shows that the pulsar's
potential drop could also be much larger than $\Phi_{\rm c}$ if
the inclination angle $\alpha=12.4^{\rm o}$ is considered, since
the assumption of alignment in RS75 is a rather strong assumption.
For the general case of oblique rotators ($\alpha \ne 0$),
\citet{xu01} proposed that the magnetic momentum of dipole
magnetic field could be expressed as $\mathbf{\vec \mu} =
\mathbf{\vec \mu}_\bot + \mathbf{\vec \mu}_{\parallel} $, where
$\mu_\bot = \mu\sin\alpha$ and $\mu_{\parallel} = \mu\cos\alpha$.
In this way, the energy-lose rate $\dot E$ consists two parts:
that of magnetic dipole radiation ($\Edot_\bot$) and that of
particle ejection due to the unipolar generator
($\Edot_{\parallel}$).
\citet{xu01} showed that the sum of these two parts are of the
same order of pure dipole radiation. Therefore, the canonical
relation $B \propto (P \dot{P})^{1/2}$ can be approximately valid.
We use the form, with a correction for $M$ and $R$, to estimate
the magnetic field,
\be B \simeq 6.4\times10^{19}(P\pdot)^{1/2}
(\frac{M}{1.4\msun})^{1/2} (\frac{R}{10 ~\mathrm{km}})^{-2}
\mathrm{~~~~G}. \label{eq:B} \ee
Consequently, the maximum potential drop would be
\be \Phimax \simeq \frac{\Omega B R^2}{2c} \cos\alpha (\sin^2\theta_2
- \sin^2\theta_1 ), \label{Phi} \ee
where $\theta_1 = \alpha - \theta$ (if $\alpha - \theta >0$) or
$0$ (if $\alpha - \theta \leq 0$) and $\theta_2 = \alpha+\theta$
(if $\alpha+\theta < 90^\circ$)
or $90^\circ$ (if $\alpha+\theta \geq 90^\circ$).
In this case, $\sin\theta = r_\mathrm{pc}/R =
({R}/{R_\mathrm{LOFL}})^{1/2}$, where $R_\mathrm{LOFL}$ is the
maximum distance of the points in the last open field line (note
$R_\mathrm{LOFL} \neq R_\mathrm{LC}$, but of the same order).

\begin{figure}
\plotone{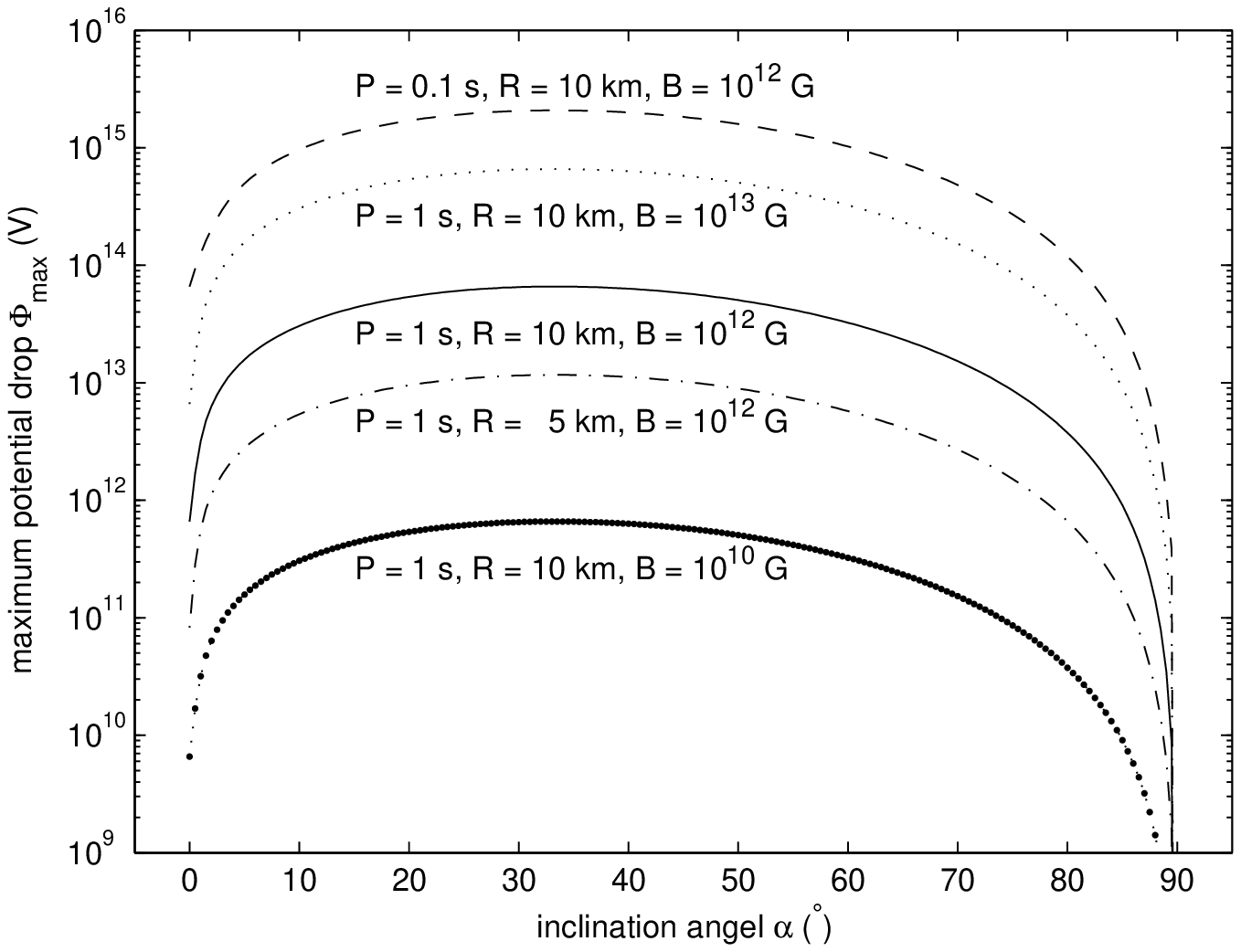}
\caption{
Maximum potential drops in the last open field line region
for different spin period ($P$), radius ($R$), and magnetic field strength ($B$).
\label{fig:univ}}
\end{figure}

\begin{figure}
\plotone{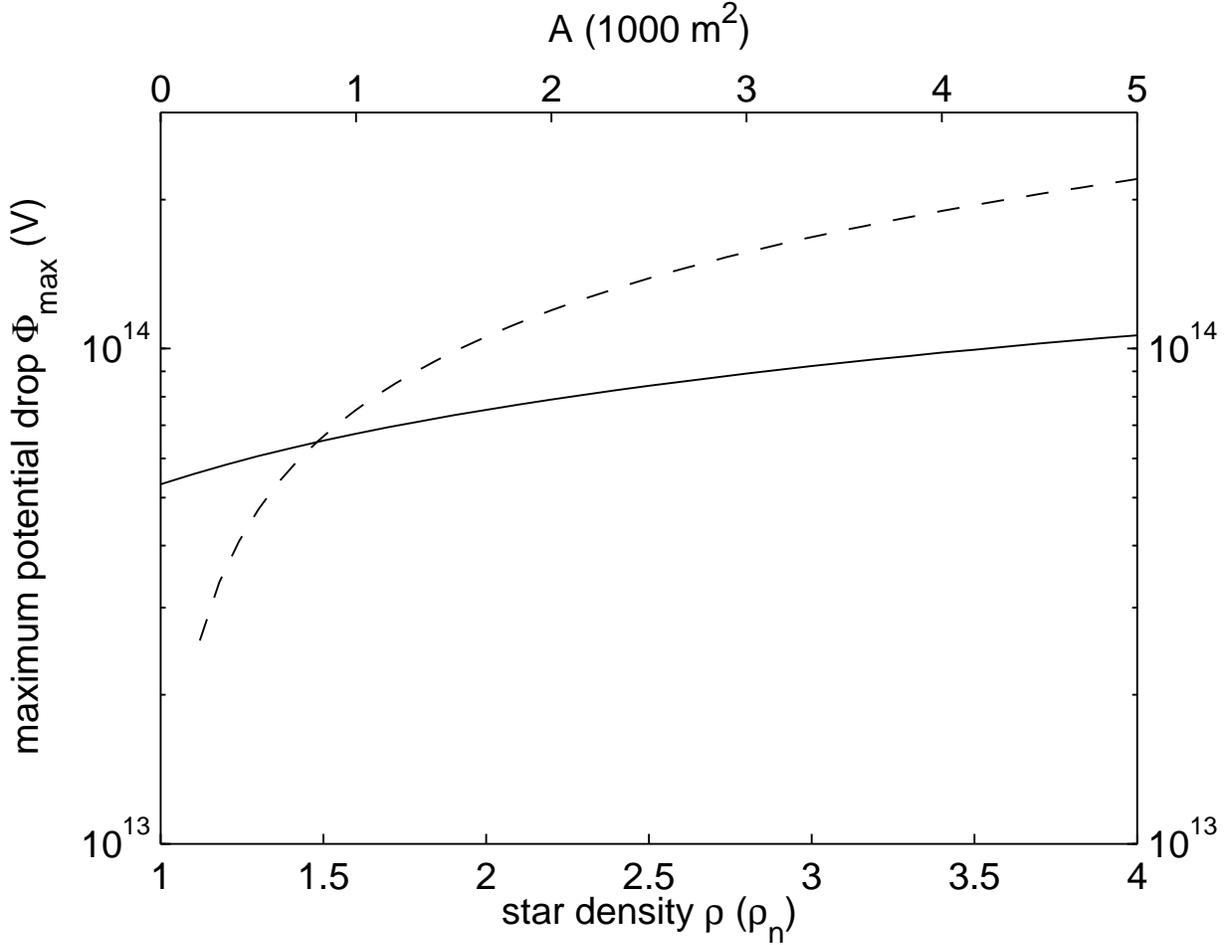}
\caption{Maximum potential drops $\Phimax$ versus stellar density $\rho$ (solid line)
and polar cap area $A$ (dashed line) of PSR B0943+10. We use
inclination angle $\alpha = 12.4^\circ$ here. The area, $A$, is in
the range of $0.3\times10^3$--$5\times10^3$ m$^2$ which is the
1-$\sigma$ uncertainty range of the observation from \citet{zhang05}.
The value $A = 10^3 \mathrm{~m^2}$ is used for calculating
the solid line; $\rho = 2\rho_\mathrm{n}$ is fixed for the
dashed line. It is evident that $\Phimax$ is well above $10^{12}$ V
($\sim\Phi_\mathrm{c}$).
\label{fig:star}}
\end{figure}

In Fig. \ref{fig:univ}, we plot $\Phimax$ versus $\alpha$ for
different $\rho$, $R$, and $B$ and find that the maximum potential
drop $\Phimax$ varies with inclination angle $\alpha$. For most
part in the range 0--90 degrees, $\Phimax$ is nearly two
magnitudes larger than that when $\alpha = 0$.
The inclination angle of PSR B0943+10 is 12.4 degrees \citep{Lyne88}.
The other parameters with large uncertainty are $\rho$ and $A$.
The maximum potential drop $\Phimax$ is a function of these two.
In Fig. \ref{fig:star}, we show the potential drop versus $\rho$
and $A$. One can see that the potential drop could be well above
$\Phi_\mathrm{c}$ ($\sim 10^{12}$ V).

\section{Conclusion and discussion}

The pulsar PSR B0943+10 could be a low-mass bare quark star of
radius $\sim 2.6$ km and mass $\sim 0.02$ $\msun$. The polar cap
area can fit the observed value of $\sim 10^3 \mathrm{~m^2}$. We
have taken into account the effect of inclination angle, which is
rarely considered previously. The maximum potential drop for PSR
B0943+10 with $\alpha =12.4^{\rm o}$ could be $10^{13\sim 14}$ V
(Fig. 2). As a consequence, the pulsar's magnetosphere would still
be active (i.e., the star would be above the death-line) and thus
be radio loud.
Our model is only a modification of RS75 model: the only
difference is that the central star is a bare quark star rather
than a normal neutron star. General features of RS75 keep in our
model (e.g., subpulse drifting would still happens). However,
there might be two advantages in quark star model: the binding
energy problem and the necessary of $\mathbf{ \Omega\cdot B<0 }$
in RS75 model do not exist anymore since the binding energy of
quarks and electrons here is nearly infinity \citep{xu01}.

Could Planck-like emission radiate from the polar caps of bare
quark stars?
This is a real question in the study of astrophysical quark matter
with low temperature but high baryon density, which could be
separated into the following two. (i) Could the emissivity of bare
strange stars be high enough to produce thermal photon due to high
plasma frequency? (ii) Can a bare strange star keep a hot polar
cap due to high thermal conductivity?
Actually, because of the uncertainties of knowledge about this
kind of quark matter, several speculations are proposed. Besides
color superconductivity \citep[CSC; e.g.,][]{csc}, \citet{xu03b}
suggested alternatively that quark matter with low temperature
should be in a solid state according to the different
manifestations of pulsar-like stars.
CSC occurs if quarks are in a condensation in momentum space;
quark clusters may form if quarks are in a condensation in
position space due to strong interaction, and quark matter could
be in a solid state if the temperature is lower than the
interaction energy between quark clusters.
The plasma frequency derived from fluid quark matter should
not be applicable in this solid quark star model.
However, the idea that quark matter with low temperature could be
in solid state is only a conjecture since no strong constrain is
given by experiments or observations. Though QCD is believed to be
the theory that would describe quark matter, it is now far from
telling us how quark matter behaves. As a result, we also could
not know certainly whether quark matter is in a solid or a fluid
phase according to today's knowledge. Nevertheless, there could be
some hints for solid quark matter, e.g., free precession
\citep[e.g.,][]{zhu06}, glitch \citep[e.g.,][]{zhou04}, and
Planckian emission \citep[e.g.,][]{xu02, Drake02,
Burwitz03,xu03b}.

The observed X-ray spectrum can also be fitted by a power law
\citep{zhang05}, which could indicate emission from magnetosphere.
If only part of X-rays is the thermal emission from the polar cap,
the derived cap radius becomes smaller, and our model still works.
It needs further observations to determine whether the emission is
thermal or non-thermal or has both components.
At the same time, thermal emission does not mean blackbody
emission. Although the ultra deep observations of RX J1856-3754
\citep[from both {\it Chandra} and {\it XMM-Newton},
e.g.,][]{Burwitz03} showed a very high-quality Planck-like
spectrum, which may hint a quark surface, blackbody emission is
only an assumption for simplicity since we do not know the
emissivity exactly.
For PSR B0943+10, if its thermal emissivity is $\sim 60$ times
smaller than that of blackbody, a star of $\sim10$ km would also
be possible, but it is serious problem to obtain theoretically
such a low emissivity for neutron star atmospheres with ions and
electrons.
Could the emissivity of quark star surface be significantly small?
If this is the possible, the star's radius would be larger than
$\sim 2.6$ km.

There are two correction factors for the polar cap area. One is
the projection factor. Actually, the observation shows only a
projected area \citep{zhang05}, but the real polar cap area could
be larger since we are not situated at the direction of the
magnetic axis. The polar cap area should be divided by a
projection factor. It can be estimated as $[\cos(\zeta+\alpha) +
\cos(\zeta-\alpha)]/2 \simeq 0.93 $. The modification is smaller
than 10\%. Hence the conclusions presented would not change. The
other factor comes from the shape of the polar cap, because it is
not an exact circle when $\alpha\neq 0$. Since $\alpha$
($=12.4^\circ$) is small, this effect is also negligible
\citep{Qiao04}.

We use $\theta_1$ and $\theta_2$ as the boundary angle (see the
definitions under eq.[\ref{Phi}]). In this way we get the maximum
potential drop. The effective potential drop depends on how the
charge current flows, which can hardly be derived from first
principles. If we use $\alpha$ instead of $\theta_1$, as in the
conventional cases, the maximum potential drop will be about half
the value. However, this does not affect our conclusion
significantly because the effective potential drop should be a
few$\times10^{12}\mathrm{~V}$ \citep[which is just
above $\Phi_\mathrm{c}$, e.g., RS75;][]{Usov95}.
Once the gap potential drop increases to $\ga \Phi_{\rm c}$,
sparks forms, the gap discharges, and the potential decreases. As
particles flow out, the potential drop increases again. Thus, if
maximum potential is $> \Phi_{\rm c}$, the RS75-type vacuum gap
model works.

The $B \propto (P \dot{P})^{1/2}$ approximation is derived by
assuming pure magnetic dipole radiation ($\alpha = 90^\circ$),
where the effect of $\alpha$ is not included.
Actually, there could be two different braking mechanisms for
isolated radio pulsars to spin down: the Poynting flux of the
magnetodipole radiation and the relativistic particle ejection due
to the unipolar generator.
The observed braking index
\citep[between 1.4 and 2.9,][and references therein]{Livingstone06}
could be naturally understood if these two mechanisms are combined
\citep{xu01,Contopoulos06}.
These different torques result in spindown-powers of a same order.
\citet{Spitkovsky06} has shown that the pulsar luminosity has
a weak dependence on $\alpha$: $L \propto 1+\sin^2\alpha$. 
He uses a simple dipole magnetic field configuration which is also
assumed in RS75 and in our model. His result could also be
applicable for quark star because a quark star may gain a dipole
magnetic field by spontaneous magnetization \citep{xu05}. The only
difference is that a quark star could be of low mass. The polar
magnetic field can be express as (note that \citet{Spitkovsky06}
calculated for the magnetic field at the magnetic equator while we
do for the polar magnetic field),
\be B = 5.2 \times 10^{19} (P\pdot)^{1/2}
(\frac{M}{1.4\msun})^{1/2} (\frac{R}{10 ~\mathrm{km}})^{-2}
(1+\sin^2\alpha)^{-1/2}~~~~{\rm G}.
\label{eq:Balpha}
\ee
The dependence on $\alpha$ is weak: $B$ only differs by a
factor of $\sqrt{2}$ at most. Thus the $B \propto (P
\dot{P})^{1/2}$ approximation is applicable.
Meanwhile, the value from  equation (\ref{eq:Balpha})
is quite close to equation (\ref{eq:B}).

Though \citet{Deshpande99, Deshpande01} has addressed a measured
drifting period of sparks on the cap of PSR B0943+10, it is still
a complex and controversial problem to detect the real drifting
rate. The the observed drifting period may possibly not be the
real one because of the aliasing effect \citep[e.g.,][]{Gil03,
Esamdin05}: several different drifting rates can fit well the same
observation. Therefore, a relatively faster drifting rate might
not be a serious problem in this bare quark star model as well as
in the RS75 model.

\acknowledgments

We are grateful to very helpful communications with an anonymous
referee.
We also would like to thank Dr. Bing Zhang for his comments and
suggestions and to acknowledge various stimulating discussions in
the pulsar group of Peking university. This work is supported by
National Nature Science Foundation of China (10573002) and by the
Key Grant Project of Chinese Ministry of Education (305001).

\clearpage

\clearpage

\end{document}